\documentstyle[psfig]{mn}

\newif\ifAMStwofonts

\ifoldfss
  \ifCUPmtlplainloaded \else
    \NewTextAlphabet{textbfit} {cmbxti10} {}
    \NewTextAlphabet{textbfss} {cmssbx10} {}
    \NewMathAlphabet{mathbfit} {cmbxti10} {} 
    \NewMathAlphabet{mathbfss} {cmssbx10} {} 
  \fi
  \ifAMStwofonts
    \ifCUPmtlplainloaded \else
      \NewSymbolFont{upmath} {eurm10}
      \NewSymbolFont{AMSa} {msam10}
      \NewMathSymbol{\upi}     {0}{upmath}{19}
      \NewMathSymbol{\umu}     {0}{upmath}{16}
      \NewMathSymbol{\upartial}{0}{upmath}{40}
      \NewMathSymbol{\leqslant}{3}{AMSa}{36}
      \NewMathSymbol{\geqslant}{3}{AMSa}{3E}

      \let\leq=\leqslant 
      \let\geq=\geqslant 
    \fi
  \fi
\fi 

\ifnfssone
  \newmathalphabet{\mathit}
  \addtoversion{normal}{\mathit}{cmr}{m}{it}
  \addtoversion{bold}{\mathit}{cmr}{bx}{it}
  \newmathalphabet{\mathbfit} 
  \addtoversion{normal}{\mathbfit}{cmr}{bx}{it}
  \addtoversion{bold}{\mathbfit}{cmr}{bx}{it}
  \newmathalphabet{\mathbfss} 
  \addtoversion{normal}{\mathbfss}{cmss}{bx}{n}
  \addtoversion{bold}{\mathbfss}{cmss}{bx}{n}
  \ifAMStwofonts
    \ifCUPmtlplainloaded \else
      \UseAMStwoboldmath
      \makeatletter
      \new@mathgroup\upmath@group
      \define@mathgroup\mv@normal\upmath@group{eur}{m}{n}
      \define@mathgroup\mv@bold\upmath@group{eur}{b}{n}
      \edef\UPM{\hexnumber\upmath@group}
      \new@mathgroup\amsa@group
      \define@mathgroup\mv@normal\amsa@group{msa}{m}{n}
      \define@mathgroup\mv@bold\amsa@group{msa}{m}{n}
      \edef\AMSa{\hexnumber\amsa@group}
      \makeatother
      \mathchardef\upi="0\UPM19
      \mathchardef\umu="0\UPM16
      \mathchardef\upartial="0\UPM40
      \mathchardef\leqslant="3\AMSa36
      \mathchardef\geqslant="3\AMSa3E

      \let\leq=\leqslant 
      \let\geq=\geqslant 
    \fi
  \fi
\fi 

\ifnfsstwo
  \DeclareMathAlphabet{\mathbfit}{OT1}{cmr}{bx}{it}
  \SetMathAlphabet\mathbfit{bold}{OT1}{cmr}{bx}{it}
  \DeclareMathAlphabet{\mathbfss}{OT1}{cmss}{bx}{n}
  \SetMathAlphabet\mathbfss{bold}{OT1}{cmss}{bx}{n}
  \ifAMStwofonts
    \ifCUPmtlplainloaded \else
      \DeclareSymbolFont{UPM}{U}{eur}{m}{n}
      \SetSymbolFont{UPM}{bold}{U}{eur}{b}{n}
      \DeclareSymbolFont{AMSa}{U}{msa}{m}{n}
      \DeclareMathSymbol{\upi}{0}{UPM}{"19}
      \DeclareMathSymbol{\umu}{0}{UPM}{"16}
      \DeclareMathSymbol{\upartial}{0}{UPM}{"40}
      \DeclareMathSymbol{\leqslant}{3}{AMSa}{"36}
      \DeclareMathSymbol{\geqslant}{3}{AMSa}{"3E}

      \let\leq=\leqslant 
      \let\geq=\geqslant 
    \fi
  \fi
\fi 

\ifCUPmtlplainloaded \else
  \ifAMStwofonts \else 
    \def\upi{\pi}
    \def\umu{\mu}
    \def\upartial{\partial}
  \fi
\fi

\title[The density of extremely red objects around AGN]
{The density of extremely red objects around high-$z$ radio-loud 
AGN\thanks{Based on observations made at the European Southern
Observatory (La Silla, Chile).}
} 
\author[A. Cimatti et al.]
       {A. Cimatti,$^1$ D. Villani,$^2$ L. Pozzetti,$^1$ S. di Serego
	   Alighieri,$^1$\\
        $^1$Osservatorio Astrofisico di Arcetri, Largo E. Fermi 5, I-50125,
		Firenze, Italy\\
		$^2$Dipartimento di Astronomia, Universit\`a di Firenze, Largo
		E. Fermi 5, I-50125, Firenze, Italy}
\date{Accepted ... 
      Received ...;
      in original form ...}

\pagerange{\pageref{firstpage}--\pageref{lastpage}}
\pubyear{....}

\begin{document}

\maketitle

\label{firstpage}

\begin{abstract}
We present the results of a $K$-band imaging survey of 40 arcmin$^{2}$
in fields around 14 radio-loud AGN (6 radio galaxies and 8 quasars)
with $z>1.5$. The survey, 80\% complete to $K<19.2$ and complemented 
by $R$-band imaging, aimed at investigating whether extremely 
red objects (EROs) are present in excess around high-$z$ AGN, and to 
study the environment of $z>1.5$ radio galaxies and radio-loud quasars. 
At $18<K<19$ the differential galaxy counts in our fields suggest a 
systematic excess over the general field counts. At $K<19.2$, we 
find an excess of galaxies with $R-K>6$ compared to the general field. 
Consistently, we also find that the $R-K$ colour distribution of all 
the galaxies in the AGN fields are significantly redder than the colour 
distribution of the field galaxies. On the other hand, the distribution
of the $R-K$ colours is undistinguishable from that of galaxies taken 
from literature fields around radio-loud quasars at $1<z<2$. We discuss 
the main implications of our findings and we compare the possible 
scenarios which could explain our results.
\end{abstract}


\section{Introduction}

One of the main open questions of galaxy evolution is how and when
the present-day massive ellipticals formed. On one hand, hierarchical
models predict that such galaxies formed at moderate redshifts through 
merging and accretion characterized by modest star formation rates 
(e.g. White \& Frenk 1991). On the other hand, the so called monolithic 
scenario predicts that the present-day ellipticals formed at very high 
redshifts through a single, intense and rapid burst of star formation 
followed by a passive and pure luminosity evolution (PLE) of the stellar 
population (e.g. Eggen et al. 1962; Larson 1975). 

From the observational point of view, the above problems can be 
investigated by searching for ``old'', passively evolved spheroidal 
galaxies at high-$z$. Such galaxies would have the distinctive signature 
of very red colours due to the absence of star formation and to the 
strong K-correction effect. 

A population of extremely red objects (hereafter EROs) was indeed
discovered with the combination of optical and near-IR imaging (e.g. 
Elston et al. 1988; McCarthy et al. 1992; Hu \& Ridgway 1994). EROs are 
found in empty sky fields (e.g. Cohen et al. 1999; Thompson et al. 1999;
Yan et al. 2000), 
in the vicinity of high-$z$ AGN (e.g. McCarthy et al. 1992; Hu \& Ridgway 
1994) and as counterparts of faint X-ray (e.g. Newsam et al. 1997;
Lehmann et al. 1999) and weak radio sources (e.g. Spinrad et al. 1997). 
Their nature is still poorly known because their faintness hampers 
spectroscopic observations and because their colours are ambiguous. In 
fact, such colours can be due not only to an old stellar population, but 
also to strong dust reddening in a star-forming or active galaxy. 

A clear example of this ambiguity is provided by HR10 ($I-K\sim 6.5$): its 
spectral energy distribution (SED) is consistent with that of an old 
elliptical at $z\approx2.4$ (Hu \& Ridgway 1994), but optical, near-IR 
and submm observations proved that HR10 is a dusty starburst galaxy at 
$z=1.44$ (Graham \& Dey 1996; Cimatti et al. 1998; Dey et al. 1999). 
Recent results showed that HR10 is not the only ERO detected in the
submm, and that $\geq$10\% of the faint submm population down to
850$\mu$m fluxes of a few mJy could be EROs (Smail et al. 1999).

On the other hand, further observations suggested that also old (ages of
$>2$ Gyr) passively evolving galaxies populate the ERO class. The best 
examples are provided by LBDS 53W091 ($R-K=5.8$; $z=1.55$; Dunlop et al. 
1996; Spinrad et al. 1997), ERO CL 0939+4713B ($R-K=7$; $z\sim 1.6$; Soifer 
et al. 1999), and by most of the EROs at $z\sim 1.3$ around the QSO 
1213-0017 (Liu et al. 2000). Also near-IR spectroscopy confirmed that 
both dusty and old galaxies contribute to the ERO population (Cimatti 
et al. 1999).

Understanding the nature and deriving the abundance of EROs is 
important to shed light on the controversial issue of the deficit  
of high-$z$ elliptical galaxies: according to some results, the number of 
galaxies with the red colours expected for high-$z$ passively evolved 
spheroidals is lower compared to the predictions of passive luminosity 
evolution (e.g. Kauffmann, Charlot \& White 1996; Zepf 1997; Franceschini 
et al. 1998; Barger et al. 1999). However, other works did not confirm 
the existence of such a deficit up to $z\sim 2$ (e.g. Totani \& Yoshii 
1997; Benitez et al. 1999; Broadhurst \& Bowens 1999; Schade et al. 1999).

As a first step to understand the nature of EROs, we started an 
imaging survey program aimed at selecting complete samples of
such galaxies both in ``empty'' fields and around high-$z$ radio-loud 
AGN. In this paper, we present the results of our survey around 
radio-loud AGN at $z>1.5$. The main motivations of such a survey are 
to investigate whether EROs are more numerous around high-$z$ AGN 
(as suspected in previous works; e.g. McCarthy et al. 1992; Dey et al. 
1995), to select old passively evolving galaxies at $z>1$ by their
optical/near-IR colours, to study the environment of high-$z$ 
radio-loud AGN (e.g. Hall, Green \& Cohen 1998; Hall \& Green 1998 and 
references therein), and 
to search for galaxy cluster candidates at $z>1.5$. We recall that 
the most distant spectroscopically confirmed cluster known to date 
has $z=1.27$ (Stanford et al. 1997; Rosati et al. 1999). Other 
works selected cluster candidates around quasars at $z>1.3$
and suggested a significant heterogeneity of the cluster galaxies,
including both passively evolving old ellipticals as well as
younger and dusty systems (e.g. Hall \& Green 1998; Liu et al. 2000).
Throughout this paper we assume $H_0=50$ kms$^{-1}$ Mpc$^{-1}$, 
$\Omega_0=1$ and $\Omega_{\Lambda}=0$ unless otherwise stated.

\section{Definition of the sample}

The survey presented here is based on $R$- and $K^{\prime}$-band 
imaging. We observed totally 14 fields: 6 around radio galaxies (RGs) 
taken from the MRC sample selected at 408 MHz (McCarthy et al. 1997, 
Kapahi et al. 1998), 8 around radio-loud quasars (RLQs) taken from the 
PKS sample selected at 2.7 GHz (Wright \& Ostrupcek 1990). We note that
MRC1017-220 is a broad line radio galaxy (Kapahi et al. 1998). All the
targeted AGN have $1.5<z<2.0$, with the exception of the quasars PKS
1351-018 ($z=3.71$) and PKS 1556-245 ($z=2.82$). The radio powers
at rest-frame 5 GHz range from $2-5 \times 10^{27}$ WHz$^{-1}$ for
RGs to $4-30 \times 10^{27}$ WHz$^{-1}$ for RLQs. Table 1 lists the 
relevant information about the sample. The fields were selected 
according to the redshifts of the AGN, to their Galactic latitude 
($>20^{\circ}$) and to their good observability during the telescope 
runs.

We have used the predictions of the Bruzual \& Charlot (1999)
evolutionary synthesis models to define a colour selection 
criterion capable to select a complete sample of EROs at high-$z$.
In particular, we used instantaneous burst models (also called
simple stellar population models, SSP) with different
redshifts of formation ($z_{f}=2,3,4,5,6$) to describe old
spheroidal galaxies. In addition, we also considered models
with exponential time-scale for the star formation rate,
$SFR \propto exp(-t/\tau)$, with $\tau=0.1,0.3$ Gyr for $\Omega=
1,0.1$ respectively. Such models correspond to cases with low 
residual star formation at $z<2$ (i.e. $<1$ M$_{\odot}$yr$^{-1}$ 
for a galaxy with mass M$_{gal}=10^{11}$ M$_{\odot}$), and they
are capable to reproduce the colours and the spectra of local
ellipticals, as well as the faint galaxy optical colour and 
redshift distributions (Pozzetti, Bruzual \& Zamorani 1996).
A Salpeter IMF ($0.1<m<125$ M$_{\odot}$) and solar metallicity 
have been adopted. Models with Scalo IMF or with supersolar
metallicities reach even redder colours at high-$z$. 
Optical to near-IR $R-K$ colours derived from the adopted 
evolutionary models are shown in Fig. 1 for two cosmologies
($H_0=50$ kms$^{-1}$ Mpc$^{-1}$, $\Omega=0.1,1$).

The colour selection threshold adopted in our survey is $R-K>6$.
This allows us to select old galaxies at $z>1.2$ formed at $z_{f}>3$ 
in both cosmologies. It is relevant to note that for $\tau>$0.3 Gyr 
and $z_{f}<3$, colours $R-K>6$ are never reached. In other words, colours
$R-K>6$ select old galaxies which had a short episode of star formation 
in early cosmological epochs. For instance, assuming no dust 
extinction, a galaxy at $z\approx$1.5 with $R-K>6$ should have 
$z_{f}>3$, an age $>$2 Gyr ($\Omega_0=1.0$) or $>$3 Gyr ($\Omega_0=0.1$) 
and $SFR<<1$ M$_{\odot}$yr$^{-1}$. 

\begin{figure}
\centerline{\psfig{figure=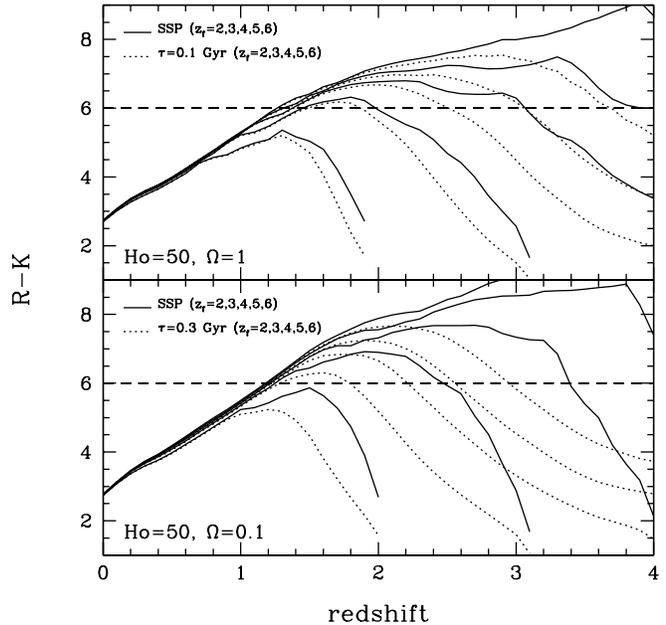,height=10cm}}
\caption[]{The $R-K$ colours predicted for different star formation
modes, formation redshifts ($z_f=2,3,4,5,6$ from left to right) and 
cosmologies (assuming no dust extinction). The solid lines represent 
single instantaneous burst models (SSP). The dotted lines show models 
with star formation rate ($SFR$) exponentially decreasing with time, 
$SFR \propto exp(-t/\tau)$, with $\tau=0.1$ Gyr and $\tau=0.3$ Gyr in 
the top and bottom panel respectively.
}
\end{figure}

\begin{table*}
\label{obs}
\tabcolsep 0.08truecm 
\caption{The observed sample}
\begin{tabular}{llcccccccccc} 
\hline
  & & & & & & & & & & & \\ 
Object & Type & z & P$_{5GHz}$ & $\alpha$ & $A_{B}$ & Seeing($K$) & t($K$) & Area($KR$) & Area($K$) & $K_{comp}$ & $N(R-K>6)$ \\
       &      &   & $10^{28}$WHz$^{-1}$ & & mag     & arcsec      & sec    & arcmin$^{2}$ & arcmin$^{2}$ & & \\ 
  & & & & & & & & & & & \\ \hline
MRC1017-220&RG&1.768&0.49&-0.95&0.23&1.1&4680&2.49&3.08&19.0&3\\ 
MRC1040-285&RG&1.630&0.17&-0.67&0.14&1.2&5228&2.88&2.88&19.1&2\\
MRC1048-272&RG&1.558&0.44&-0.73&0.27&1.3&3780&2.95&3.31&18.9&3\\
MRC1131-269&RG&1.711&0.18&-0.71&0.24&1.0&4200&2.82&2.82&18.9&1\\
PKS1143-245&RLQ&1.950&1.30&-0.18&0.22&1.0&3920&2.65&2.84&18.9&0\\
PKS1148-001&RLQ&1.976&2.90&-0.48&0.03&1.0&4200&2.73&3.00&18.9&0\\
PKS1148-171&RLQ&1.751&0.50&-0.30&0.04&1.2&4320&2.88&3.49&18.9&0(1)$^{\ast}$\\
MRC1217-276&RG&1.899&0.16&-1.20&0.31&1.2&5180&2.75&3.32&18.9&0\\
MRC1259-200&RG&1.580&0.39&-1.05&0.24&1.0&4320&3.66&3.66&18.9&0\\
PKS1351-018&RLQ&3.707&1.90&+0.04&0.11&1.1&4320&2.35&2.79&18.9&2\\
PKS1354-152&RLQ&1.890&1.80&-0.50&0.24&1.0&4200&3.33&3.33&19.0&0\\
PKS1403-085&RLQ&1.763&0.63&-0.33&0.06&1.0&4200&2.83&2.83&19.0&0\\
PKS1504-164&RLQ&1.790&0.40&-0.68&0.42&1.0&3744&2.76&2.76&18.9&0(2)$^{\ast\ast}$\\
PKS1556-245&RLQ&2.813&1.50&-0.40&0.51&1.0&4200&2.90&3.25&19.0&0\\
  & & & & & & & & & & & \\
  & & & & & & & & & & & \\ \hline
\end{tabular}

\footnotesize
Columns: (2) RG: radio galaxy; (3) RLQ: radio-loud quasar; (4) P$_{5GHz}$: 
radio power at rest-frame 5 GHz; (5) $\alpha$: radio spectral index between
2.7 GHz and 5.0 GHz ($S_{\nu} \propto \nu^{\alpha}$) according to the 
NASA/IPAC Extragalactic Database (NED); (6) $A_{B}$: Galactic extinction 
in $B$-band; (7) t($K$): total integration time in $K$-band; (8) 
Area($KR$): common area between the $K$- and the $R$-band images; (9)
Area($K$): area covered by the $K$-band image; (10) $K_{comp}$: $K$-band 
magnitude corresponding to the 100\% completeness level; (11) $N(R-K>6)$: 
number of objects with $K<19.2$ and $R-K>6$. $^{\ast}$: the field of PKS 
1148-171 contains one object with $K<19.2$ and $R-K>5.75$ ($3\sigma$ lower 
limit). $^{\ast\ast}$: the field of PKS 1504-164 contains two objects with 
$K<19.2$ and $R-K>5.35$ and $R-K>5.48$ ($3\sigma$ lower limits).

\end{table*}

\section{Observations, data reduction and photometry}

Near-infrared imaging was done on 1997 April 1-3 with the ESO/MPI 2.2m 
telescope with the IRAC2B camera (Moorwood et al. 1992) equipped with 
a 256$\times$256 HgCdTe array (0.506$^{\prime \prime}$/pixel). We
used the $K^{\prime}$ filter in order to reduce the thermal noise.
The sky conditions were photometric and the seeing was around 
1.0$^{\prime \prime}$. The observations were done taking a number of
background-limited images (typically 10-14) with the telescope moved 
$10^{\prime \prime}$ between each image. In each telescope position,
the exposures were typically 120 seconds long (e.g. 12 images each
with an exposure time of 10 seconds). The data reduction was performed 
using the IRAF reduction package and using the method outlined by 
Villani \& di Serego Alighieri (1999). Photometric calibration was 
obtained observing standard stars from the Carter \& Meadows (1995) 
sample. The typical night-to-night scatter in the zero-points was 
around 0.03 magnitudes. The conversion from $K^{\prime}$ to $K$ 
magnitudes in the SAAO-Carter system was done following Lidman \& 
Storm (1995) and Lidman (1997, private communication). 

Optical imaging was done in service mode from February 1997 to April 
1997 with the ESO NTT 3.5m telescope equipped with the SUSI camera 
(1024$\times$1024 CCD, 0.130$^{\prime \prime}$/pixel). The conditions 
were photometric and the seeing ranged from 0.6$^{\prime \prime}$ 
to 1.2$^{\prime \prime}$. The observations were done taking 5-6 images 
(each with a duration of 600 seconds) with the telescope moved 10
arcsec between each exposure. The total integration time in
each field was 3000 seconds, with the exceptions of MRC1040-285 and
PKS1351-018 where it was 3600 seconds. The photometric calibration was 
obtained using the Landolt (1992) standard stars. The typical night-
to-night scatter of the zero-points was around 0.05 mag. The correction 
for the atmospheric 
extinction was done using the $V$-band coefficients ($a_{V}$) derived by 
the Swiss Telescope at La Silla during the nights of our NTT observations 
(see Burki et al. 1995), and converting them to $R$-band extinction adopting 
$a_{R}=0.55 a_{V}$ according to the average atmospheric extinction curve
of La Silla. The data were reduced using IRAF. A ``super'' 
flat-field image was obtained by taking the median of 60 images of the 
observed AGN fields and rejecting the discrepant pixels. After subtracting 
the bias and dividing for the flat-field image, the individual frames 
of each field were coadded in order to obtain the final images.

Photometry was done using the SExtractor image 
analysis package (Bertin \& Arnouts 1996). The detection threshold
was set to $3\sigma$ of the background intensity in a contiguous
area of $\approx$1 seeing FWHM. We first identified and extracted the 
objects in the $K^{\prime}$-band images, and then we cross-correlated 
them with the $R$-band images. The depth of the $R$-band images is 
rather homogeneous, providing a $S/N \approx 3$ for objects with 
$R \approx 25$ (with a 3$^{\prime \prime}$ diameter photometric 
aperture).  The magnitudes were measured in 3$^{\prime \prime}$ 
diameter apertures and converted to 6$^{\prime \prime}$ magnitudes 
by applying the aperture corrections estimated from field stars.
Such corrections are typically in the range of 0.08-0.15 and 
0.04-0.10 magnitudes in $K^{\prime}$- and $R$- bands respectively. 
Photometric errors were 
estimated according to the poissonian noise from the objects and 
from the background. Finally, the magnitudes were corrected for 
Galactic extinction using the $A_{B}$ values derived from the 
extinction maps of Burnstein \& Heiles (1982) and adopting 
$A_{R}=0.56 A_{B}$ and $A_{K}=0.07 A_{B}$ (see Table 1).

The completeness in each $K^{\prime}$ field was estimated
by constructing a background image for each field and adding random
simulated objects with Moffat profiles consistent with the observed
seeing. The simulated objects were then detected with SExtractor as 
a function of magnitude. The completeness is rather homogeneous among
all the fields. The global completeness of the whole $K^{\prime}$-
selected sample is 100\% for $K=19.0 \pm 0.1$ and it decreases to 
about 80\% for $K=19.2 \pm 0.1$. The common sky area covered by
$R$- and $K$-band imaging is 40.0 arcmin$^2$. At $K<19.2$, only
3 objects are undetected in $R$-band (see Figure 3).

\section{$K$-band galaxy counts}

The main uncertainty in galaxy counts comes from the star--galaxy
separation. Because of the poor pixel sampling in the IRAC2B
images, the morphological classifier used by SExtractor is more 
reliable at moderately bright magnitudes. Such classifier, 
called ``stellarity index'' ($S$), is by definition equal to 1.0 for 
stars and to 0.0 for galaxies. We found that $S>0.9$ provided a reliable 
classification of stellar objects at $K<$17.5. We then decided to 
statistically subtract 
the star counts using the method outlined by Saracco et al. (1997): 
taking advantage of the good seeing and better sampling of the
SUSI images, we extracted a subsample of objects that were reliably 
classified as stellar in the $R$-band. In order to assess the fraction
of objects that are misclassified in the near-IR images, we computed 
the ratio $R_s = N_{s,K}/N_{s,R}$, where $N_{s,K}$ and $N_{s,R}$ are
the number of objects that SExtractor classifies as stars (i.e. with
$S>0.9$) respectively in the IRAC2B and in the
SUSI images. The ratio $R_s$ was computed as a function of the
$K$-band magnitude over a range 15.0$<K<$19.0. As expected, $R_s\sim1$
for 15.0$<K<$17.0, indicating that SExtractor correctly classifies 
the stellar objects at moderately bright $K$-band magnitudes.
However, the classification efficiency drops to $R_s\sim0.5$ for 
17.0$<K<$19.0, showing that at fainter $K$-band magnitudes SExtractor 
underestimates the number of stars in IRAC2B images.

\begin{figure}
\centerline{\psfig{figure=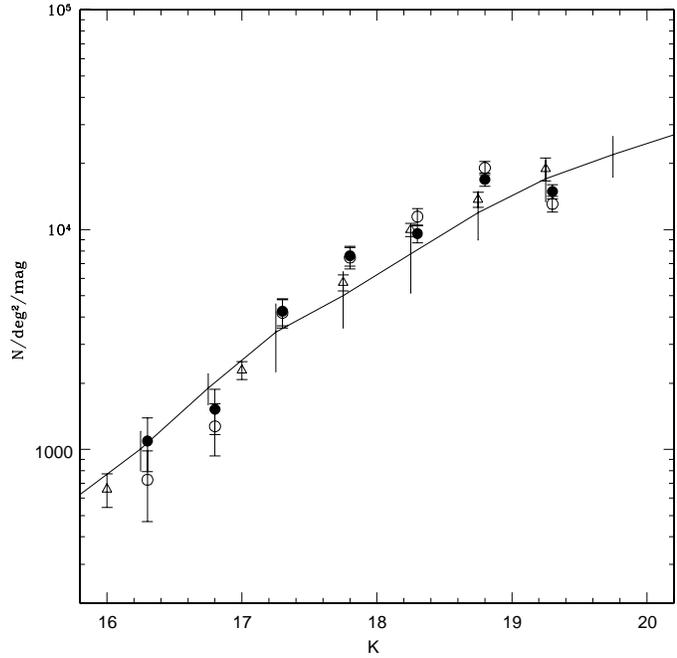,height=10cm}}
\caption[]{The observed differential galaxy counts (black circles).
The error bars on our counts indicate the field-to-field rms.
The two vertical dashed lines show respectively the 100\% and 80\% 
completeness limits in $K$-band. The counts at $K=19.3$ are 
$\approx$70\% incomplete. The open circles show the counts derived
by selecting objects with $R$-band stellarity $<0.9$ (see text).
The continuum line represents the literature-averaged counts taken 
from Hall \& Green (1998) with their rms. The triangles are the 
counts taken from the Minezaki et al. (1998) survey.}
\end{figure}

The galaxy counts were then corrected adopting the $R_s$ values 
computed for each $K$-band magnitude bin. We excluded from the 
counts the objects lying at $<3^{\prime \prime}$ from the edges 
of the images in order to avoid photometric uncertainties due
to edge effects. This implied a reduction of the useful total 
area of the 14 fields from 46.4 arcmin$^{2}$ to 43.4 arcmin$^{2}$.

Table 2 and Figure 2 show the galaxy counts in $K$-band as derived from 
the observed fields compared with the literature averaged counts of
Hall \& Green (1998) and the counts of the wide-field (181 arcmin$^{2}$) 
survey of Minezaki et al. (1998). For $K<17.5$, our counts are consistent 
with the average literature counts. For $17.5<K<19$ the counts seem 
to have a systematic excess of a factor of $\approx$1.2-1.5 respect to 
average literature counts. 
For $K>19$ the counts are incomplete in most of the fields (see Tab 2). 
In order to provide an estimate of the field-to-field variations of
the counts, we do not show their poissonian uncertainties, but 
the their rms derived from the different fields. 

We also tried an alternative approach deriving the $K$-band counts
by selecting objects with $R$-band ``stellarity index'' $<0.9$ from
the total area in common between the $K$- and the $R$-band images
(40 arcmin$^2$). Within the poissonian uncertainties, we obtained 
results consistent with the previous method, thus independently 
confirming the existence of an excess of counts for $K>17.5$
(see Figure 2). 

\begin{table}
\label{obs}
\tabcolsep 0.08truecm 
\caption{The $K$-band galaxy counts}
\begin{tabular}{lccr} 
\hline
  & & & \\ 
$K$ & Completeness & $n$ & $\Delta n$ \\
  & & & \\
16.3& 100\% &  1092 &  300\\
16.8& 100\% &  1523 &  355\\
17.3& 100\% &  4250 &  596\\
17.8& 100\% &  7603 &  793\\
18.3& 100\% &  9587 &  890\\
18.8& 100\% & 16918 & 1183\\
19.3& $\sim$70\%  & 14907 & 1112\\
  & & & \\
  & & & \\ \hline
\end{tabular}

\footnotesize
$n$: number density of galaxies deg$^{-2}$ mag$^{-1}$; 
$\Delta n$: poissonian uncertainty on $n$.

\end{table}

\section{The morphologies of EROs}

Figure 3 shows the colour magnitude diagram for all the objects 
(stars+galaxies) extracted from the 40 arcmin$^2$ of our RG and 
RLQ fields in common between the $K$- and the $R$-band imaging. 
At $K<18.9$ and at $K<19.2$, 8 and 11 objects with $R-K>6$ are selected 
respectively. The faintest EROs have $25.5<R<26.0$ and they are
detected at $2\sigma - 3\sigma$ significance level in the $R$-band.
In addition, we found three objects which are clearly detected
in the $K$-band at $K<19.2$, but which are invisible in the 
$R$-band. Their $3\sigma$ lower limits on the colours are in the 
range of $R-K>5.3-5.8$ (see Tab. 1 and Fig. 3).

\begin{figure}
\centerline{\psfig{figure=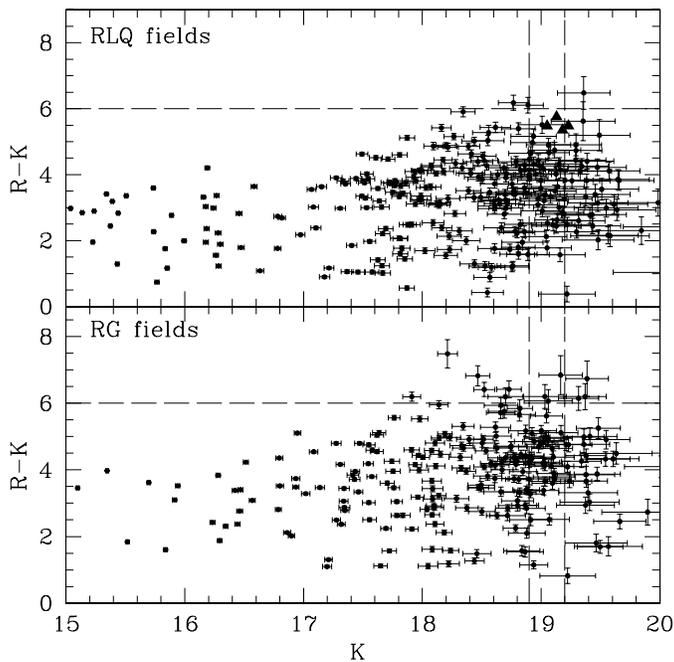,height=10cm}}
\caption[]{The colour-magnitude diagrams for all the objects in
the radio galaxy and radio-loud quasar fields. The horizontal dashed
lines indicates the $R-K>6$ selection threshold. The two vertical
dashed lines show respectively the 100\% and 80\% completeness
limits in $K$-band. The filled triangles indicate the objects undetected 
in $R$-band and their $3\sigma$ lower limits in the $R-K$ colours.}
\end{figure}

As discussed by Thompson et al. (1999) and Cuby et al. (1999),
low mass stars and brown dwarfs can contribute to the population 
of very red objects. A visual inspection of the detected
objects with $R-K>6$ suggested a clear non-stellar morphology for most
of them. However, in order to establish more quantitatively if our 
sample is contaminated by stars, we analyzed the morphology of the $R-K>6$ 
objects by means of their ``stellarity'' indices in $K$- and in $R$-band, 
$S_K$ and $S_R$. 

The stellarities in $R$-band are in the range of 
$0.15<S_R<0.57$, and clustered around $S_R$=0.35, indicating that
the morphologies are rather ambiguous in the optical. This is 
not surprising because the objects with $R-K>6$ are so faint in the optical 
($R\approx 25$) that the morphological classifier becomes less reliable
at such low flux levels. Thus, the morphological analysis can be better 
done in the $K$-band. In the $K<18.9$ sample, one object is unresolved 
($S_K=0.96$), one is ambiguous ($S_K=0.44$), and six are clearly 
non-stellar ($S_K<0.15$). In the $K<19.2$ sample, one object is 
unresolved, three are ambiguous ($0.4<S_K<0.5$), and seven are clearly 
non-stellar ($S_K<0.15$). We noticed that the FWHM of the ambiguous 
objects is $\approx 1.2-1.8$ times larger than the seeing, suggesting 
that such objects may be compact galaxies rather than stars. Among 
the three objects undetected in $R$, one is clearly non-stellar 
($S_K=0.18$) and the other two are compact (both have $S_K=0.85$). 

In order to have more clues, we also used other two 
morphological parameters provided by SExtractor: the elongation and 
the ellipticity, which should be respectively equal to 1.0 and 0.0 
for perfect stellar objects. In our $K$-band images, we found that, 
for reliable stellar objects ($S_K>0.95$) their average 
values are respectively $1.10\pm0.05$ and $0.08\pm0.04$, whereas
the objects with $R-K>6$ have values in the range of 1.21-2.39 and 
0.14-0.58 respectively. The only exception is the object with $S_K=0.96$ 
which, consistently with its stellarity index, has an elongation of 
1.15 and an ellipticity of 0.13. The analysis of the elongation and 
the ellipticity favours the detected objects with $R-K>6$ being mostly 
non-stellar objects. 

\section{The surface density of EROs}

\subsection{Our AGN fields}

The total surface densities of EROs with $R-K>6$ averaged over the 
14 observed fields derived for the $K<18.9$ and the $K<19.2$ samples 
are respectively 0.20$\pm$0.07 and 0.28$\pm$0.08 objects arcmin$^{-2}$. 
However, we notice that there are strong variations of the density of 
objects with $R-K>6$ from field to field, reaching values as
high as $\sim$1 arcmin$^{-2}$ in some of the observed fields
(see Tab. 1). The errors indicated in this section are 1$\sigma$ 
poissonian uncertainties.

The average densities decrease respectively to 0.18$\pm$0.07 and to 
0.25$\pm$0.08 objects arcmin$^{-2}$ if we exclude the unresolved object. 
Thus, the average density of {\it galaxies} with $R-K>6$ is in the range 
of 0.15-0.18($\pm$0.06-0.07) objects arcmin$^{-2}$ for the $K<18.9$ sample, 
depending on the exclusion or the inclusion of the object with ambiguous 
morphology. For the $K<19.2$ sample, the galaxy density is 0.18-0.25
($\pm$0.07-0.08) arcmin$^{-2}$, but it could increase to 0.33$\pm$0.09 
arcmin$^{-2}$ in case the 3 objects undetected in $R$-band were galaxies 
with $R-K>6$. If we exclude from our sample the densest fields of MRC 
1017-220 and MRC 1040-285, the density of objects with $R-K>6$ and 
$K<19.2$ decreases to $0.17\pm0.07$ arcmin$^{-2}$.

\subsection{Hall \& Green (1998) RLQ fields}

The average surface density of EROs derived in our fields has been compared 
to that of the Hall \& Green (1998) sample (30 fields around RLQs
at $1<z<2$ covering a total area of 213.7 arcmin$^{2}$). First of
all, the $r$-band magnitudes of Hall \& Green (1998) have been 
transformed to $R$ using their relation $R=r-0.322$, and their
$K$-band magnitudes have been scaled to our $K$ photometric system 
using the conversions of Carter (1993) adopting $J-K=3$, 
consistently with the extreme $J-K$ colors of EROs (e.g. Hu \& Ridgway
1994). However, we note that the results do not change if
we adopt a bluer color such as $J-K=1.5$. From the Hall \& Green (1998)
sample, we estimate that the densities of galaxies with $R-K>6$ are 
0.21$\pm$0.03 arcmin$^{-2}$ and 0.29$\pm$0.04 arcmin$^{-2}$ at $K<18.9$ 
and $K<19.2$ respectively. Such densities agree very well with those
estimated in our AGN fields. 

\subsection{Field surveys}

Our derived surface densities have been also compared to the results 
of surveys of the general field (i.e. not around high-$z$ AGN)
based on $R$- and $K$-band photometry. In this respect, the main
reference survey is that of of Thompson et al. (1999) because it 
has the widest field coverage available to date (154 arcmin$^{2}$), 
and it is therefore the least affected by the field-to-field variations 
present in other surveys with areas about an order of magnitude smaller. 
Since Thompson et al. (1999) made use of non-standard filters ($R_{CADIS}$ 
and $K^{\prime}$), we derived colour and magnitude thresholds consistent 
with our photometric system adopting the relations $K=K^{\prime}-0.22(H-K)$ 
(Thompson, private communication; see also Wainscoat \& Cowie 1991) and 
$H-K=1$, and $R=R_{CADIS}-0.1$ (valid for $B-R=2$, Huang \& Thompson, 
private communication). We adopted $H-K=1$ because the observed
colors of most EROs are in the range of $0.7<H-K<1.2$ (Thompson,
private communication; see also Hu \& Ridgway 1994; Spinrad et al. 1997;
Stanford et al. 1997; Soifer et al. 1999; Stiavelli et al. 1999; Yan 
et al. 2000). 

Excluding the 2 EROs in the Thompson et al. sample which are 
spectroscopically confirmed stars, we found that the densities 
of EROs are 0.05$\pm$0.02 arcmin$^{-2}$ (8 objects) and 0.11$\pm$0.03 
arcmin$^{-2}$ (17 objects) at $K<18.9$ and $K<19.2$ respectively. 
We mention here that Daddi et al. (2000), who performed a survey 
for EROs covering a field of 701 arcmin$^{2}$ and 447 arcmin$^{2}$ 
to $K<18.8$ and $K<19.2$ respectively, found surface densities 
of EROs with $R-K>6$ fully consistent with those of Thompson et 
al. (1999).

We also derived the density of galaxies with $R-K>6$ from the
combination of other
three field surveys: McLeod et al. (1995) (14.7 arcmin$^{2}$), Cohen 
et al. (1999) (14.6 arcmin$^{2}$), and from the two control fields of
Hall \& Green (1998) (18.8 arcmin$^{2}$). The $K$-band magnitudes of 
the literature samples have been scaled to our photometric system 
using the transformations of Carter (1993) and assuming an average 
colour of $J-K=3$, but the results do not change if we assume 
$J-K=1.5$. The $r$-band magnitudes of Hall \& Green (1998) have 
been transformed to $R$ as described in section 6.2. Only three 
galaxies with $R-K>6$ and $K<19.2$ are present in the joint sample 
of such surveys over a total area of 48.1 arcmin$^{2}$. The derived 
surface density is 0.06$\pm$0.04 arcmin$^{-2}$ which is consistent 
with that of Thompson et al. (1999). 

The above comparisons imply that there is a overdensity of galaxies 
with $R-K>6$ in fields around radio-loud AGN at $z>1.5$ compared to 
the general field. At $K<18.9$ and $K<19.2$ the average density
in our AGN fields is respectively a factor of 4 and 2.5 higher than
the density in the general field derived by Thompson et al. (1999).
In particular, if we take the field densities of Thompson et al. (1999),
we expect 2 and 4.4 EROs with respectively $K<18.9$ and $K<19.2$ in our 
40 arcmin$^2$ survey. The Poisson probability of our observing 
8 and 11 EROs with $K<18.9$ and $K<19.2$ is 0.086\% and 0.37\%,
implying an overdensity at 99.9\% and 99.6\% confidence levels
respectively. 
However, it is also important 
to notice that strong field-to-field variations are present in our 
survey, and that in some of the fields the density of EROs is about 
10 times higher than in the general field (see Tab.1). Our results
suggest that the strong field-to-field variations are due to presence 
of high-$z$ clusters in some of the observed fields. It is relevant
to mention here that other observations found evidence of ERO
overdensities around high-$z$ AGN (e.g. McCarthy et al. 1992; Dey et
al. 1995; Hall \& Green 1998; Clements 2000; Liu et al. 2000).

\section{Color distributions}

\subsection{Comparison with field galaxies}

In order to investigate the colour properties of all the galaxies 
detected in the 14 AGN fields, we compared their $R-K$ colours with 
those of 245 field galaxies selected from the samples of McLeod 
et al. (1995), Cohen et al. (1999), and Hall \& Green (1998). 
First of all, in order to check the validity of our field sample, we 
compared the average $R-K$ colours of our literature sample with those 
of the Thompson et al. (1999) sample. For $16.8\leq K < 17.8$ and 
$17.8\leq K < 18.8$, we found average $R-K$ colours ($3.60\pm0.11$ 
and $3.83\pm0.09$ respectively) which are consistent with those 
derived by Thompson et al. (1999) in their survey ($3.69\pm0.05$ 
and $3.83\pm0.04$ respectively).

Figure 4 shows the colour distributions of the galaxies selected from
the literature field sample and from our AGN fields. Since both 
distributions contain lower limits on the $R-K$ colours, we used the 
ASURV Rev 1.2 statistical software package (see LaValley, Isobe \& 
Feigelson 1992). 

\begin{figure}
\centerline{\psfig{figure=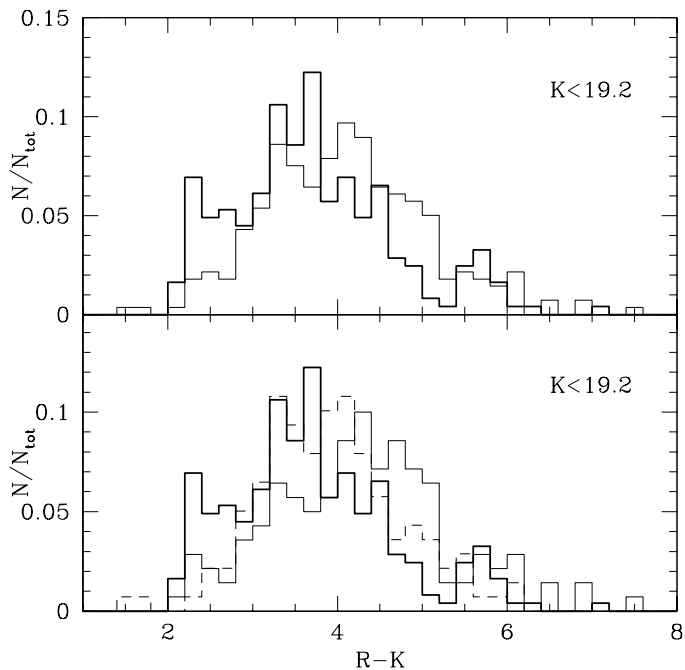,height=10cm}}
\caption[]{The colour distributions of galaxies. Upper panel:
the thin line represents our total sample (radio galaxies
and quasars), whereas the thick line shows the distribution
of colours of field galaxies. Lower panel: the thin continuum
and dashed lines show the colour distributions of the galaxies 
respectively in the radio galaxy and in the quasar fields.
Thick line: same as in the upper panel.}
\end{figure}

As already suggested by Figure 4, we found that the galaxies with $K<19.2$ 
selected in AGN fields are redder than those selected from the general 
field. In particular, all the statistical tests available in ASURV
show consistently that the two distributions are different at $\geq$99.99\%
($\geq 4 \sigma$) significance level. Such result does not change 
significantly if we split our total sample into the RG and the RLQ 
sub-samples: the RG sample is redder than the literature field sample 
at $\geq$99.99\% significance, whereas the RLQ sample is redder at 
$99.0$\% significance level.

If we exclude from this analysis the four fields with the highest 
densities of EROs (i.e. MRC1017-220, MRC1040-285, MRC1048-272 and 
PKS1351-018), the result does not change significantly: the galaxies 
in the 
AGN fields continue to be redder than the field galaxies at 99.95\% 
confidence level (about 3.5$\sigma$). This suggests that red
colour excess is not due to particularly overdense fields,
but to the general population of galaxies in all the AGN fields.
Such results strengthens the result on the overdensity of EROs
with $R-K>6$ discussed in the previous section.

Finally, we found that the mean $R-K$ colours estimated with the 
Kaplan-Meier estimator for the literature, (RG+RLQ), RG, and RLQ samples 
are respectively: 3.717, 4.108, 4.289, and 3.921.

\subsection{Comparison with literature AGN fields}

It is also important to compare the colour distributions of
the galaxies in our AGN field sample with other samples of
AGN fields taken from the literature. To this purpose, we used the data 
collected by Hall \& Green (1998) observing 30 fields containing 
radio-loud quasars: 
10 at redshifts $1.0<z<1.4$, and 20 at $1.4<z<2.0$.  
The 5 GHz radio powers of such quasars span the range $26.0<$log 
P$<28.2$ WHz$^{-1}$. From the Hall \& Green (1998) sample we selected
278 and 899 galaxies with $K<19.2$ respectively in the quasar
redshift ranges $1.0<z<1.4$ and $1.4<z<2.0$. Figure 5 already
shows that
the colour distributions of the galaxies in our fields and in the 
Hall \& Green (1998) quasar fields look very similar. This is confirmed 
by the statistical tests which show that the colours of the galaxies 
in our (RG+RLQ), RG and RLQ fields are not significantly different 
from those of the Hall \& Green (1998) fields, irrespective of the redshift 
range. The results of sections 7.1 and 7.2 do not significantly 
change if the two RLQs with $z>2$ are excluded (PKS1351-018 and 
PKS1556-245).

\subsubsection{RGs vs. RLQs}

The results discussed in section 7.1 show that our RG and RLQ 
sub-samples have different average colours ($R-K$=4.289 vs. $R-K$=3.921
respectively).
Such a difference is also confirmed by the comparison of their colour 
distributions, for which the statistical tests indicate that the two 
distributions are different at 99.9\% confidence level. However, because 
of the 
limited statistics of our survey (only 6 RG fields vs. 8 RLQ fields), we
have compared the colour distributions of the RG fields with the
colour distribution of large Hall \& Green (1998) quasar sample
(30 fields at $1<z<2$). The statistical tests indicated that
the two distributions are not signficantly different (i.e.
they are different at $\approx 1.7 \sigma$ level). The result
does not significantly change if we add our RLQ sample to the 
Hall \& Green (1998)
sample, and if subsamples in different redshift bins are considered
(e.g. $1.0<z<1.4$ and $1.4<z<2.0$). Again, the results do not change
if we exclude PKS1351-018 and PKS1556-245 from the statistical analysis. 
Our conclusion is that the colour difference that we found between our 
RG and RLQ sub-samples is probably due to the limited statistics of our
survey rather than to a real diversity.

\begin{figure}
\centerline{\psfig{figure=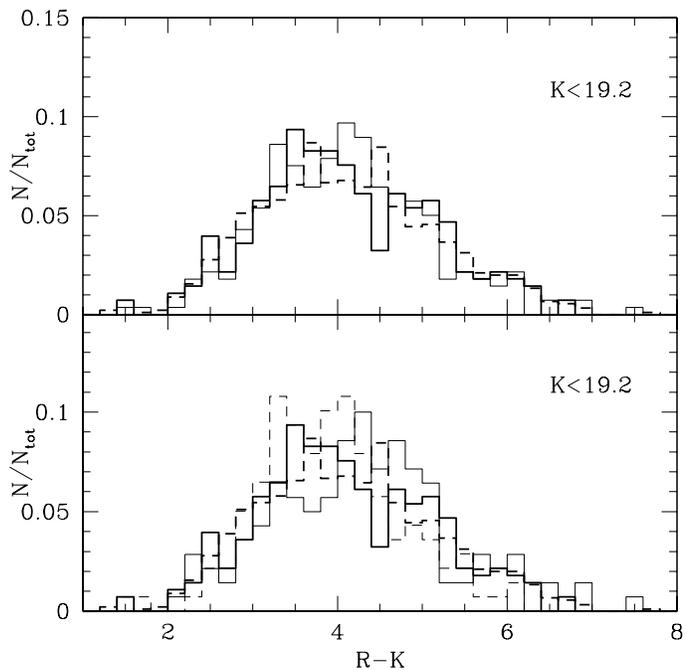,height=10cm}}
\caption[]{The colour distributions of galaxies. Upper panel:
the thin line represents our total sample (radio galaxies
and quasars); the thick continuum and dashed lines show the 
colour distributions of galaxies in fields around radio-loud quasars
at $1.0<z<1.4$ and $1.4<z<2.0$ respectively (from Hall \& Green 
1998). Lower panel: the thin continuum and dashed lines show the 
colour distributions of the galaxies respectively in the radio galaxy 
and in the quasar fields. Thick lines: same as in the upper panel.}
\end{figure}

\subsection{Searching for radial trends}

Despite the rather shallow $K$-band depth of our sample and the 
limited sky area covered by each $K$-band image, we investigated
the possible existence of correlations of the radially-averaged 
colour and the number 
density of galaxies with the radial projected distance from the AGN. 
This is useful to test if there is a tendency of the reddest galaxies
to lie closer to the AGN, and to investigate whether the AGN tend to
be located in denser environments, as found by Hall \& Green (1998).
Figure 6 shows the radial dependences of the average $R-K$ and number
density of galaxies with $K<19.2$ in 5 arcsec wide bins. The results 
are strongly influenced by the limited statistics, and no significant 
trends are visible. However, we notice that the average $R-K$ seems
to be higher going closer to the AGN. Should such trend be confirmed
by deeper imaging, this would mean that the reddest galaxies are
preferentially located close to the AGN. We also note that in 
the bins closer to the AGN the galaxy number density seems to be 
systematically higher than the density of galaxies in the general 
field. 
Finally, we mention that our sample is not deep enough to allow us an 
evaluation of the cluster richnesses using estimators such as the 
Hill \& Lilly (1991) $N_{0.5}$ statistic.

\begin{figure}
\centerline{\psfig{figure=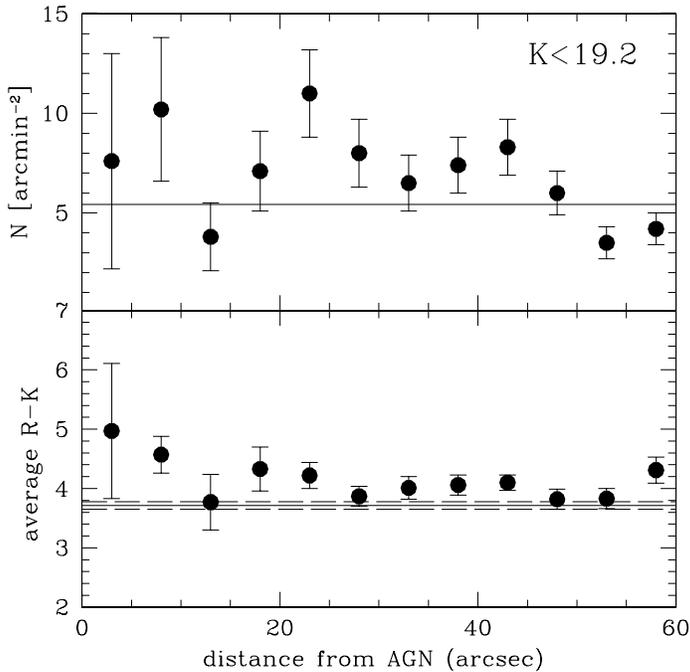,height=10cm}}
\caption[]{Upper panel: the cumulative average density of galaxies
with $K<19.2$ as a function of the radial distance from the AGN.
The continuum line represents the cumulative number density of
field galaxies to $K<19.2$ derived from the literature averaged 
counts of Hall \& Green (1998). Lower panel: the average $R-K$
colours of galaxies with $K<19.2$ as a function of the radial 
distance from the AGN. The continuum and dashed lines represent
the average $R-K$ colour and its $\pm1 \sigma$ dispersion as 
derived from the literature field galaxies (see section 7.1).} 
\end{figure}

\section{Main implications}

The main finding of our survey is the discovery of an excess of 
red galaxies in fields around radio galaxies and radio-loud 
quasars at $z>1.5$. Our results are in good agreement with
the ones of the Hall \& Green survey around RLQs at $1<z<2$.
Being our survey rather shallow ($K<19$) and based only on 
two-band photometry, deep multi-colour photometry and spectroscopy 
are needed to unveil the redshifts and the nature of the excess 
red galaxies. With the present results, we 
can simply envisage two main possible scenarios to explain our finding.

{\bf (A)} 
The excess galaxies are at the same resdhift of the AGN ($z=z_{AGN}$) 
in the field, and they represent the ``tip of the 
iceberg'' of the clusters or the groups where the AGN live. This 
possibility is generally supported by previous works which showed 
that RLQs and RGs are sometimes found in rich environments over a wide 
range of reshifts: $0.3<z<1$ (e.g. Yee \& Green 1987; Elligson, 
Yee \& Green 1991; Hill \& Lilly 1991; Deltorn et al. 1997), 
$1<z<3$ (Aragon-Salamanca et al. 1996; Le Fevre et al. 1996; Pascarelle
et al. 1996; Dickinson 1997; Yamada et al. 1998; Hall \& Green 1998; 
Pentericci et al. 1998). On the other hand, it is relevant to recall 
here that radio-quiet QSOs at most redshifts are found in lower density 
environments (e.g. Elligson et al. 1991; Teplitz et al. 1999 and 
references therein).

A first possibility is that the observed EROs are star-forming
galaxies strongly reddened by dust extinction (e.g. Cimatti
et al. 1998; Dey et al. 1999; Smail et al. 1999). In such a case, 
an excess of dusty star-forming galaxies may suggest that high-$z$ 
radio-loud AGN live in environments characterized by an intense 
star-formation activity. In this respect, it is relevant to mention
that Ivison et al. (2000) recently found an excess of submm
sources with possible ERO counterparts around the $z=3.8$ radio 
galaxy 4C 41.17, suggesting an excess of dust enshrouded star 
formation in a proto-cluster environment. Such result, if 
confirmed and generalized to other cases, would be relevant to 
strengthen a scenario where high-$z$ radio-loud AGN live 
in rich environments where strong star formation is taking place.

Alternatively, the observed EROs could be passively evolving spheroidals 
with negligible or no ongoing star formation activity, indicating that 
some of the observed AGN may live in clusters or groups where, similarly 
to lower-$z$ clusters, old and luminous spheroidals represent a 
substantial fraction of the galaxies. 

Follow-up deep imaging and spectroscopy of the most overdense
fields (e.g. MRC 1017-220 and MRC 1040-285) is under way in order
to verify the existence of clusters at $z=z_{AGN}$. Preliminarly,
near-infrared spectroscopy of two of the reddest galaxies in the 
field of MRC 1017-220 ($z=1.77$) suggested that their redshifts
are $z\sim 1.5 \pm 0.25$ and that their SEDs are consistent 
with that of $\approx$2-3 Gyr-old elliptical galaxies with no dust
extinction (Cimatti et al. 1999), strengthening the hypothesis 
that the red galaxies are probably at the same redshift of the AGN.

Assuming that the excess galaxies are ellipticals at the reshifts
of the AGN, we estimated their rest-frame $K$-band absolute magnitudes
in case of formation redshift $z_f>z_{AGN}>3$. For the cases at
$1.5<z_{AGN}<2.0$ and adopting $z_f=4$, we obtain that $M_K$ are in 
the range of -25.5$\div$-26.5 and -26$\div$-27.5 for ($\Omega=1$, 
$\tau=0.1$ Gyr) and ($\Omega=0.1$, $\tau=0.3$ Gyr) respectively. 
If compared with $L^{\ast}_{K}$ of the local luminosity function of
elliptical galaxies ($M^{\ast}_{K}=-25.16$, Marzke et al. 1998),
such luminosities suggest that the observed EROs have rest-frame
$L_{K} \sim 1.4-3.4 L^{\ast}_{K}$ and $L_{K} \sim 2.2-8.6 L^{\ast}_{K}$ 
for ($\Omega=1$, $\tau=0.1$ Gyr) and ($\Omega=0.1$, $\tau=0.3$ Gyr)
respectively. 

In the case of PKS 1351-017, if the EROs in 
its field were at $z=3.7$, their $M_K$ would be in the range of 
-26.5$\div$-27.5 and -28$\div$-29.5 for ($\Omega=1$,$\tau=0.1$ Gyr) 
and ($\Omega=0.1$, $\tau=0.3$ Gyr) respectively, implying very high 
luminosities ($L_{K} \sim 3.4-8.6 L^{\ast}_{K}$ for $\Omega=1$, $\tau=0.1$
Gyr), and probably too high ($L_{K} \sim 14-55 L^{\ast}_{K}$) in the 
case of ($\Omega=0.1$, $\tau=0.3$ Gyr). However, we notice that if 
the local $L^{\ast}_{K}$ evolves and increases at high-$z$, as 
suggested by some recent works (e.g. Glazebrook et al. 1995;
Schade et al. 1999), the luminosity of the EROs in excess over 
the local $L^{\ast}_{K}$ would be reduced.  

In this regard, it is important to emphasize that the presence 
of moderately bright galaxies with extreme red colours such 
as $R-K>6$ strongly suggests the existence a population of 
passively evolving ``old'' galaxies at $1.5<z<2$ whose colours 
can be explained only if their formation redshifts are $z_{form}>3$.
Should this result be confirmed by the redshift measurements
of the most extreme EROs, this would provide very stringent
clues on the earliest epoch of massive galaxy formation
(see for instance Spinrad et al. 1997).

{\bf (B)}
The second scenario is that the excess red galaxies are
at $z \not= z_{AGN}$ and physically unrelated to the AGN.

First of all, the observed EROs could be dust reddened
starburst galaxies. For instance, HR10 provides a clear
case of a foreground star-forming galaxy at $z=1.4$
located in the field of a quasar at $z=3.8$ (Hu \& Ridgway
1994). In this respect, the two EROs that we detected in 
the field of PKS 1351-017 may be foreground objects 
unrelated to the quasar. 

If $z > z_{AGN} >2$, extreme colours such as $R-K>6$ can be
reached only if the galaxies formed at $z_f>3$ for instantaneous
burst models or $z_f>4$ for models with $\tau=0.1$ or 0.3 Gyr
(see Fig. 1). In such a case, the detected EROs would represent
moderately old and very luminous passively evolving galaxies
at $z>2$, and they would provide important clues on the existence
of ``aged'' galaxies at high-$z$. Although the existence of such
galaxies is in principle possible, the fraction of galaxies
with $K<19$ and $z>2$ expected in galaxy formation models is 
extremely small, ranging from $\sim$1\% in PLE scenarios (e.g.
Pozzetti et al. 1996) to $\sim$0\% in hierarchical models (e.g.
Kauffmann \& Charlot 1998). This means that in our fields
we expect {\it at most} 4 galaxies with $K<19$ and $z>2$.
We notice here that at $K<19$, no galaxies with $z>2$ have 
been spectrospically identified yet in the redshift surveys 
of Cowie et al. (1996) and Cohen et al. (1999). 

The other possibility is that the red galaxies are part
of foreground clusters or structures at $z < z_{AGN}$.
Although this case is realistic, it does 
not easily explain why radio-loud AGN are preferentially
seen through foreground structures (in order to explain
the excess of EROs around radio-loud AGN). In this case, 
it is tempting to speculate whether the mass of such 
foreground structure and the geometry of the system can
produce gravitational lensing effects on the background
AGN. A similar scenario was already suggested by Hammer
\& Le Fevre (1990).

At a speculative level, we can envisage a scenario where
the light coming from the AGN is gravitationally amplified
by foreground massive structures, producing a selection
effect where the AGN that would not be selected in flux-limited
surveys are actually selected thanks to the amplification of the
flux due to the gravitational lensing. In this respect, we notice 
that the 408 MHz fluxes of some of the AGN found in the richest 
environments (e.g. MRC 1017-220, MRC 1040-285) are in the range of 
1030-1090 mJy, i.e. only $\sim$8-15\% higher than the selection 
threshold of the MRC radio survey (0.95 Jy at 408 MHz) (see McCarthy 
et al. 1997 and references therein). This means that, assuming that
the intrisic non-lensed flux of the sources is just below
0.95 Jy, a small gravitational amplification of the flux by a 
factor of $\approx$10-15\% would push these AGN above the selection 
threshold of the survey. The ability of clusters of galaxies to 
gravitationally lens extended sources is clearly demonstrated by 
the luminous arcs which are background galaxies (i.e. extended sources)
distorted and amplified by the cluster potential (e.g. Smail et al.
1995). Although in our case the geometry is different, i.e. the 
distance of the lens is not much smaller than that of the source, 
we are only considering small magnification factors which could involve 
even only one radio lobe, whose size is smaller or similar to that of 
galaxies in the optical.

To summarize, the present data do not allow us to explore in more 
details the validity of the possible scenarios. However, according
to the above discussion, we can conclude that, based on their 
colours and $K$-band magnitudes, the excess red galaxies around
the AGN are likely to be at $1<z<2$, hence favoring the scenario
where radio-loud AGN at $1.5<z<2$ live in rich environments, or 
the case where the red galaxies are at $z<z_{AGN}$.
The former case is relevant to understand the genesis of the AGN 
phenomenon, the relationships between radio-quiet and radio-loud AGN 
and for the use of radio-loud AGN as tracers of cluster and massive 
structures of galaxies at high-$z$. Follow-up optical and near-IR 
spectroscopy and submillimetre photometry of the selected EROs 
is under way in order to unveil their nature and their role in the 
framework of galaxy formation and evolution.

\section*{Acknowledgments}

We are grateful to J. Cohen and P. Hall for providing the colours and 
the magnitudes of their galaxy samples, to D. Thompson for useful 
information on the CADIS photometric system and on the $H-K$ colors 
of the CADIS EROs, to C. Lidman for the information on the IRAC2b 
photometric calibration, to D. Silva for his assistance in the use 
of the SUSI data taken in service observing, and to G. Bruzual and 
S. Charlot for providing their spectral synthesis models. We acknowledge
the anonymous referee for the useful comments. This research has made use 
of the NASA/IPAC Extragalactic Database (NED) which is operated by the 
Jet Propulsion Laboratory, California Institute of Technology, 
under contract with the National Aeronautics and Space Administration. 
L.P. acknowledges the support of a research grant from {\it Consorzio
Nazionale Astronomia e Astrofisica (CNAA)} during the development of 
this project.


\begin{thebibliography}{}
\bibitem{}Aragon-Salamanca A., Ellis R.S., O'Brien K.S. 1996, MNRAS,
281, 945
\bibitem{}Barger A.J., Cowie L.L., Trentham N., Fulton E., Hu E.M.,
Songaila A., Hall D. 1999, AJ, 117, 102
\bibitem{}Benitez N., Broadhurst T.J., Bouwens R.J., Silk J., Rosati P.
1999, ApJ, 515, L65
\bibitem{}Bertin E., Arnouts S. 1996, A\&A, 117, 393
\bibitem{}Broadhurst T.J., Bouwens R.J. 1999, ApJ, in press
(astro-ph/9903009)
\bibitem{}Bruzual G., Charlot S. 1999, private communication
\bibitem{}Burnstein D., Heiles C. 1982, AJ, 87, 1167
\bibitem{}Burki G., Rufener F., Burnet M., Richard C., Blecha A., 
Bratschi P. 1995, A\&AS, 112, 383.
\bibitem{}Carter B.S. 1993, in Precision Photometry, eds. D. Kilkenny , 
E. Lastovica, J.W. Menzies, SAAO, Cape Town, p. 100 
\bibitem{}Carter B.S., Meadows V.S. 1995, MNRAS, 276, 734
\bibitem{}Cimatti A., Andreani P., R\"ottgering H., Tilanus R. 1998, Nature,
392, 895
\bibitem{}Cimatti A., Daddi E., di Serego Alighieri S., Pozzetti L., 
Mannucci F., Renzini R., Zamorani G., Andreani P., R\"ottgering H.J.A. 
1999, A\&A, 352, L45 
\bibitem{}Clements D.L. 2000, MNRAS, in press (astro-ph/0002101)
\bibitem{}Cohen J.G., Blandford R., Hogg D.W., Pahre M.A., Shopbell P.L.,
ApJ, 512, 30
\bibitem{}Cowie L.L., Songaila A., Hu E.M., Cohen J.G. 1996, AJ, 112,
839
\bibitem{}Cuby J.-G., Saracco P., Moorwood A.F.M., D'Odorico S., Lidman C., 
Comeron, Spyromilio J. 1999, A\&A, in press (astro-ph/9907028)
\bibitem{}Daddi E., Cimatti A., Pozzetti L., Hoekstra H., R\"ottgering
H.J.A., Renzini A., Zamorani G., Mannucci F. 2000, A\&A submitted
(astro-ph/0005581)
\bibitem{}Deltorn J.M., Le Fevre O., Crampton D., Dickinson M. 1997,
ApJ, 483, L21
\bibitem{}Dey A., Spinrad H., Dickinson M. 1995, ApJ, 440, 515
\bibitem{}Dey A., Graham J.R., Ivison R.J., Smail I., Wright G.S. 1999, ApJ,
519, 610
\bibitem{}Dickinson M. 1997, in HST and the High Redshift Universe, ed.
N. Tanvir, A. Aragon-Salamanca, and J.V. Wall, (Singapore: World
Scientific, p. 207
\bibitem{}Dunlop J.S., Peacock J.A., Spinrad H., Dey A., Jimenez R.,
Stern D., Windhorst R.A. 1996, Nature, 381, 581
\bibitem{}Eggen O.J., Lynden-Bell D., Sandage A.R. 1962, 136, 748 
\bibitem{}Ellingson E., Yee H.K.C., Green R.F. 1991, ApJ, 371, 49
\bibitem{}Elston R., Rieke G.H., Rieke M. 1988, ApJ, 331, L77
\bibitem{}Franceschini A., Silva L., Fasano G., Granato L., Bressan A.,
Arnouts S., Danese L. 1998, ApJ, 506, 600
\bibitem{}Glazebrook K., Peacock J.A., Miller L., Collins C.A. 1995,
MNRAS, 275, 169
\bibitem{}Graham, J.R., Dey, A. 1996, ApJ, 471, 720
\bibitem{}Hall P.B., Green R.F., Cohen M. 1998, ApJS, 119, 1
\bibitem{}Hall P.B., Green R.F. 1998, ApJ, 507, 558
\bibitem{}Hill G., Lilly S. 1991, ApJ, 367, 1
\bibitem{}Hu E.M., Ridgway S.E. 1994, AJ, 107, 1303 
\bibitem{}Ivison R.J., Dunlop J.S., Smail I., Dey A., Liu M.C., Graham
J.R. 2000, ApJ, in press (astro-ph/0005234)
\bibitem{}Kapahi V.K., Athreya R.M., van Breugel W., McCarthy P.J.,
Subrahmanya C.R. 1998, ApJS, 118, 275
\bibitem{}Kauffmann G., Charlot S., White S.D.M 1996, MNRAS, 283, 117
\bibitem{}Kauffmann G., Charlot S. 1998, MNRAS, 297, L23 
\bibitem{}Landolt A.U. 1992, AJ, 104, 340
\bibitem{}LaValley M., Isobe T., Feigelson E.D. 1992, in Astronomical 
Data Analysis Software and Systems I, A.S.P. Conference Series, Vol. 
25, D.M. Worrall, C. Biemesderfer \& J. Barnes eds., p. 245.
\bibitem{}Le Fevre O., Deltorn J.M., Dickinson M. 1996, ApJ, 471, L11
\bibitem{}Lehmann I., Hasinger G., Schmidt M., Gunn J.E., Schneider
D.P., Giacconi R., McCaughrean M., Tr\"umper J., Zamorani G. 1999,
A\&A, in press (astro-ph/9911484)
\bibitem{}Hammer F., Le Fevre O. 1990, ApJ, 357, 38
\bibitem{}Larson R.B. 1975, MNRAS, 173, 671
\bibitem{}Lidman C., Storm J. 1995, in IRAC-2b Report on Test
Observations IV, ESO.
\bibitem{}Liu M.C., Dey A., Graham J.R., Bundy K.A., Steidel C.C.,
Adelberger K., Dickinson M.E. 2000, AJ, in press (astro-ph/0002443)
\bibitem{}Marzke R.O., Da Costa L.N., Pellegrini P.S., Willmer C.N.A.,
geller M.J. 1998, ApJ, 503, 517
\bibitem{}McCarthy P.J., Persson S.E., West S.C. 1992, ApJ, 386, 52 
\bibitem{}McCarthy P.J., Kapahi V.K., van Breugel W., Persson S.E.,
Athreya R., Subrahmanya C.R. 1996, ApJS, 107, 19 
\bibitem{}McLeod B.A., Bernstein G.M., Rieke M.J., Tollestrup E.V.,
Fazio G.G. 1995, ApJS, 96, 117
\bibitem{}Minezaki T., Yukiyasu K, Yoshii Y., Peterson B.A. 1998,
ApJ, 494, 111
\bibitem{}Moorwood A., Finger G., Biereichel P., Delabre B., van 
Dijsseldonik A., Huster G., Lizon J.-L., Meyer M., Gemperlein H., 
Moneti A. 1992, The Messenger, 69, 61
\bibitem{}Newsam A.M., McHardy L.M., Jones L.R., Mason K.O. 1997, MNRAS, 
292, 378
\bibitem{}Pascarelle S.M., Windhorst R.A., Driver S.P., Ostrander E.J.,
Keel W.C. 1996, ApJ, 456, L21
\bibitem{}Pentericci L., R\"ottgering H.J.A., Miley G.K., Carilli C.L,
Mccarthy P.J. 1997, A\&A, 326, 580
\bibitem{}Pozzetti L., Bruzual A.G., Zamorani G. 1996, MNRAS, 281, 953
\bibitem{}Rosati P., Stanford S.A., Eisenhardt P.R., Elston R., Spinrad
H., Stern D., Dey A. 1999, AJ, in press (astro-ph/9903381)
\bibitem{}Saracco P., Iovino A., Garilli B., Maccagni D., Chincarini G.
1997, AJ, 114, 887
\bibitem{}Schade D., Lilly S.J., Crampton D., Ellis R.S., Le F\`evre O.,
Hammer F., Brinchmann J., Abraham R., Colless M., Glazebrook K., Tresse
L., Broadhurst T. 1999, ApJ, in press (astro-ph/9906171)
\bibitem{}Soifer B.T., Matthews K., Neugebauer G., Armus L., Cohen J.G., Persson
S.E. 1999, AJ, in press (astro-ph/9906464)
\bibitem{}Smail I., Couch W.J., Ellis R.S., Sharples R.M. 1995, ApJ,
440, 501
\bibitem{}Smail I., Ivison R.J., Kneib J.-P., Cowie L.L., Blain A.W.,
Barger A.J., Owen F.N., Morrison G. 1999, MNRAS, 308, 1061
\bibitem{}Spinrad H., Dey A., Stern D., Dunlop J., Peacock J., Jimenez R.,
Windhorst R. 1997, ApJ, 484, 581
\bibitem{}Stanford S.A., Elston R., Eisenhardt P.R., Spinrad H.,
Stern D., Dey A. 1997, AJ, 114, 2232
\bibitem{}Stiavelli M., Treu T., Carollo M., Rosati P., Viezzer R.,
Casertano S., Dickinson M., Ferguson H., Fruchter A., Madau P., 
Martin C., Teplitz H. 1999, A\&A, 343, L25.
\bibitem{}Teplitz H.I., McLean I.S., Malkan M.A. 1999, ApJ, 520, 469
\bibitem{} Thompson D., Beckwith S.V.W., Fockenbrock R., Fried J.,
Hippelein H., Huang J.-S., von Kuhlmann, Ch. Leinert, Meisenheimer
K., Phleps S., R\"oser H.-J., Thommes E., Wolf C. 1999, ApJ, in press
\bibitem{} Totani T., Yoshii J. 1997, ApJ, 501, L177
\bibitem{}Villani D., di Serego Alighieri S. 1999, A\&AS, 135, 299
\bibitem{}Wainscoat R.J., Cowie L.L. 1992, AJ, 103, 332
\bibitem{}White S.D.M., Frenk C.S. 1991, ApJ, 379, 52
\bibitem{}Wright A.E., Ostrupcek R. (eds.) 1990, {\it PKSCAT90
database}, ANTF, Parkes, NSW, Australia.
\bibitem{}Yamada T.; Tanaka I., Aragon-Salamanca A., Kodama T., Ohta 
K., Arimoto N. 1997, ApJ, 487, L125
\bibitem{}Yan L., McCarthy P.J., Weymann R.J., Malkan M.A., Teplitz
H.I., Storrie-Lombardi L.J., Smith M., Dressler A. 2000, ApJ in press
(astro-ph/0004170)
\bibitem{}Yee H.K.C., Green R.F. 1987, ApJ, 319, 28
\bibitem{}Zepf S.E. 1997, Nature, 390, 377
\end{thebibliography}
\end{document}